\documentstyle[aps,prl,twocolumn,psfig]{revtex}
\begin{document}
\draft

\title{Energetic and spatial bonding properties from
angular distributions of ultraviolet photoelectrons: 
application to the GaAs(110) surface} 
\author{C.\ Solterbeck and W.\ Schattke, }

\address{Institut f\"ur Theoretische Physik,
Christian-Albrechts-Universit\"at Kiel, Leibnizstra{\ss}e~15,
D-24098 Kiel, Germany}

\author{J.-W.\ Zahlmann-Nowitzki, K.-U.\ Gawlik, L.\ Kipp, and
              M.\ Skibowski}

\address{Institut f\"ur Experimentalphysik,
Christian-Albrechts-Universit\"at Kiel, Leibnizstra{\ss}e~19,
D-24098 Kiel, Germany}

\author{C.\ S.\ Fadley$^{\dag \ast}$ and M.\ A.\ Van Hove$^{\ast}$}
\address{$^{\dag}$ Dept. of Physics, University of California-Davis, 
Davis, California 95616, USA}
\address{$^{\ast}$ Lawrence Berkeley Laboratory, 
University of California, Berkeley, California 94720, USA}

\maketitle

\begin{abstract}
Angle-resolved ultraviolet photoemission spectra are interpreted 
by combining the energetics and spatial properties of the contributing 
states. One-step calculations are in excellent agreement with new 
azimuthal experimental data for GaAs(110). Strong variations 
caused by the dispersion of the surface bands permit an accurate 
mapping of the electronic structure. The delocalization of the 
valence states is discussed analogous to photoelectron diffraction. 
The spatial origin of the electrons is determined, 
and found to be strongly energy dependent, with uv excitation 
probing the bonding region.\end{abstract}


Angle-resolved ultraviolet photoemission spectroscopy 
(ARUPS) is probably the most powerful single experimental 
technique for studying the valence electronic structure of solids.  
However, until recently, ARUPS has been primarily limited to 
investigating the positions in energy of valence bands along a few 
high-symmetry directions in reciprocal space. That is, the 
intensity 
of the photoemission peaks was usually not analyzed 
quantitatively, and most of the information in the full hemisphere 
above the surface was lost. One reason for this 
limitation in ARUPS 
studies is the lack of any simple rules for explaining such valence 
spectra beyond those that have been found useful for mapping 
bands in energy. The most often applied model in ARUPS is that 
of direct (wave-vector-conserving) transitions between bulk 
bands\cite{direct}, with the final state often 
being simplified to a plane wave\cite{fou2}. 
Beyond this, free-electron final states with atomic-like 
optical transitions\cite{fouat} and final-state 
scattering of electrons emerging from a localized core 
orbital\cite{diff} have been used 
to better understand the resulting angular distributions in 
photoemission. The 
interest in such angular distributions of intensity has recently been 
stimulated by the measurement of full-hemisphere 
intensity maps in 
ARUPS\cite{hemi1,hemipl,hemisc,hemidi,hemiat,nablak}, 
with the data being analyzed so far using plane-wave 
final states\cite{hemipl}, 
single scattering\cite{hemisc}, bulk bands\cite{hemidi}, 
and atomic-like transitions\cite{hemiat}, as well as the 
one-step model\cite{hemisi,hemiferm}. 
The most accurate description of valence 
ARUPS involves calculations within this one-step model 
and includes the precise optical matrix elements, full multiple 
scattering, the explicit presence of the surface potential, and the 
resulting more complex initial and final states. 
However, there has to date been no 
systematic treatment of 
the angular distribution of intensity within such a model.

In this Letter, we quantitatively investigate the influence of 
the initial state and its charge distribution on the angular 
distribution of intensity in ARUPS, using the 
one-step model. We demonstrate that the combined consideration of 
both energy positions and intensity patterns 
will be necessary for the most useful 
interpretation of angle-scanning data, and point out 
the additional information that can be derived in this way.  
For our calculations and measurements, azimuthal scans
are chosen, which allow a high accuracy in the 
visualization of the angular distribution. A generalization of 
the x-ray photoelectron diffraction (XPD) picture 
is found to lead to a more unified 
view of the electronic structure in both direct and reciprocal space. 
The GaAs(110) surface is chosen as a well-understood 
test case\cite{gaascalc1,gaascalc2,ups,gaasband} for which 
it will become evident that the photoemission patterns reveal not 
only the surface density of states (SDOS) but also provide insight 
into the charge density of the bonds. This contributes 
to a common understanding of ARUPS and XPD angular distributions.

Our interpretation of photoelectron emission 
proceeds via the construction of the photoelectron state 
$\Psi$ with energy $E_{fin}$, which can be written at the 
detector located in the direction of the polar ($\vartheta$) and 
azimuthal ($\varphi$) emission angles as
\begin{equation}
 |\Psi\rangle = \sqrt{D} \; 
                    G(E_{fin},\vartheta, \varphi) \;
                     ({\bf A \cdot p }) \; |\Psi_{in}\rangle \; ,
 \label{eq:psi_val}
\end{equation}
where $D$ is the SDOS, $G$ is the propagator, 
${\bf A}$ is the vector potential, and ${\bf p}$ is 
the momentum operator\cite{diff}. The initial state, depending 
on the energy $E_{in}$ and  
${\bf k}_{\parallel}$, is split into the factor of the SDOS $D$ 
and the wave function $\Psi_{in}$. This formula is used for 
the following discussion of the electronic and spatial structure. 
Other influences like selection rules, the density of the final 
state, or resonances of direct volume transitions are not 
explicitly apparent, but they are all included in 
Eq. (\ref{eq:psi_val}) and in a one-step calculation for which 
Eq. (\ref{eq:psi_val}) is transformed to a golden rule formula.
There is principally no difference in the validity of applying
the one-step model to both ARUPS or XPD.
In Eq. (\ref{eq:psi_val}), $D(E_{in},{\bf k}_{\parallel})$ 
contributes to the band mapping, and, as noted above, is often 
the only thing considered in most analyses to derive experimental 
band structures. 
$G$ describes the electron scattering from the 
point of excitation 
through the crystal to the detector. In XPD this scattered 
electron can be simply thought as being 
emitted from a localized initial state; 
here the general case arises 
to treat emission from delocalized valence states. 
The source of the scattered electron is no longer the vicinity of 
a core but the entire initial state wave function. 
The wave amplitude $\langle {\bf r} |\Psi_{in}\rangle$ 
in Eq. (\ref{eq:psi_val}) gives the local emissivity, depending 
on the spatial position, initial energy, parallel momentum, and, 
together with the radiation polarization, the angular momentum composition.

As a test of this interpretation, we investigate azimuthal 
distributions of the photocurrent above GaAs(110) in terms of the 
electronic and geometric structure of the participating states. 
The calculation of the photocurrent within the one-step model 
proceeds along the lines described elsewhere\cite{gaascalc1,hemisi}.
The spectra were measured with unpolarized HeI$_{\alpha}$ radiation 
($h\nu = 21.22$ eV) with the energy resolutions set to $130$ meV and,
for testing purposes, to $35$ meV, which gave negligible differences.
The scans were taken with an angle resolution of $0.25^{\circ}$ by 
use of a $180^{\circ}$ spherical electron analyzer, which is movable 
around two independent axes. These degrees of freedom permit taking 
angle scans without moving the sample and allow studying the 
influence of the incoming light and separating such effects from 
other processes.

\begin{figure}
\centerline{\psfig{figure=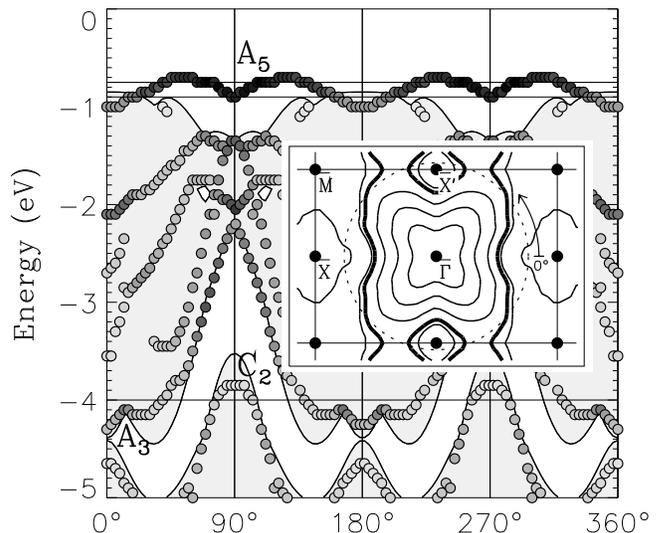}}
\caption{
Calculated surface band structure with the peaks (circles) of the SDOS 
along the circular path for $k_{\parallel} = 0.6$ {\AA}$^{-1}$. 
Strong peaks are dark. 
The projected bulk band structure is also shown (shaded). 
The inset shows the
A$_{5}$ band in a contour plot with equidistant energy levels 
and the level for $-0.75$ eV (dark).
The band disperses in the Brillouin zone 
from ${\overline{\Gamma}}$ down to a local minimum in 
${\overline{X'}}$ and its global minimum in ${\overline{X}}$.
The dotted circle describes the ${\bf k_{\parallel}}$ path 
studied here. 
All energies are referred to the valence band maximum.
}
\label{fig:1}
\end{figure}

Two cases are studied here: At $-0.75$ eV there is a 
dangling-bond state (A$_{5}$), that is a surface state lying 
outside the 
projected bulk bands, and at $-4.0$ eV there are two resonances, 
the Ga back-bond state (C$_{2}$) and the A$_{3}$ state. To 
get a broad view off the high symmetry points, we consider a 
circle in the Brillouin zone with a radius of 
$k_{\parallel} = 0.6$ {\AA}$^{-1}$, corresponding to the measurement angles 
at both energies. The surface band structure 
along this circle is shown in Fig.\ \ref{fig:1} together with the 
projected bulk band structure. The energy contour plot in the 
inset of Fig.\ \ref{fig:1} exhibits the symmetry of the 
A$_{5}$ state; this surface has only a mirror plane, but as an 
additional 
symmetry, an inversion symmetry occurs in the electronic structure 
due to Kramer's degeneracy $(E({\bf k}) = E({\bf -k}))$.

\begin{figure}
\centerline{\psfig{figure=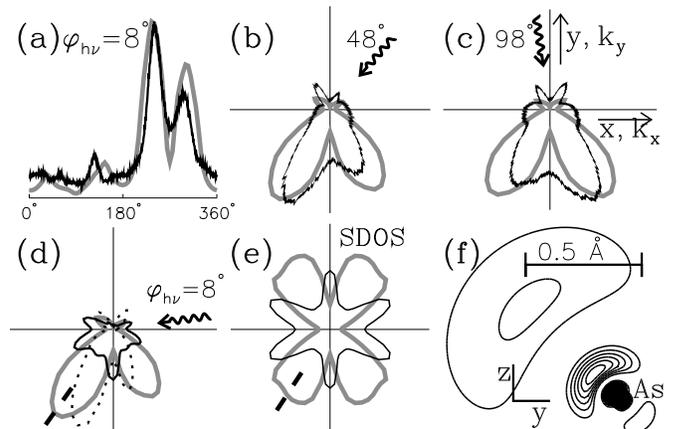}}
\caption{
Photocurrent at $k_{\parallel} = 0.6$ {\AA}$^{-1}$ 
($\vartheta = 17.3^{\circ}$), 
SDOS, and charge density of the A$_{5}$ band: 
(a) to (c) current measured at $-0.9$ eV (black) and calculated 
 at $-0.75$ eV (grey) for light incident at azimuthal angles of
 $98^{\circ}$, $228^{\circ}$, and $8^{\circ}$, respectively; 
(d) calculated current for $98^{\circ}$ incidence 
 at $-0.75$ eV (grey), at $-0.9$ eV 
 (solid), and at $-0.75$ eV with a free final state (dotted); 
(e) SDOS at $-0.75$ eV (grey)
 and at $-0.9$ eV (black); 
(f) contour plot of the charge density in a $yz$ plane at 
 $-0.75$ eV for ${\bf k_{\parallel}}$ belonging
 to the lower left lobe as indicated by the dashed lines in (d) and (e).
In the polar plots (b) to (e), the radius represents the intensity of  
 the current and the value of the SDOS. 
 Accordingly, the azimuthal angle is referred to the 
 emission direction and to the ${\bf k_{\parallel}}$ vector. The
 orientation of the axes is indicated in (c).
}
\label{fig:2}
\end{figure}

For a detailed discussion of the electronic and geometric influences
on the photocurrent, the effect of the incident light on these 
spectra must be known first. With a fixed polar 
angle of incidence and the use of unpolarized light, an asymmetry 
remains only due to the azimuthal 
direction $\varphi_{h\nu}$ of the light. Here, 
enhancements and azimuthal peak shifts are weak, as shown in 
Figs. \ref{fig:2} (a)--(c). 
These influences can be discriminated from the strong differences in the 
current between positive and negative $y$ directions.

\begin{figure}
\centerline{\psfig{figure=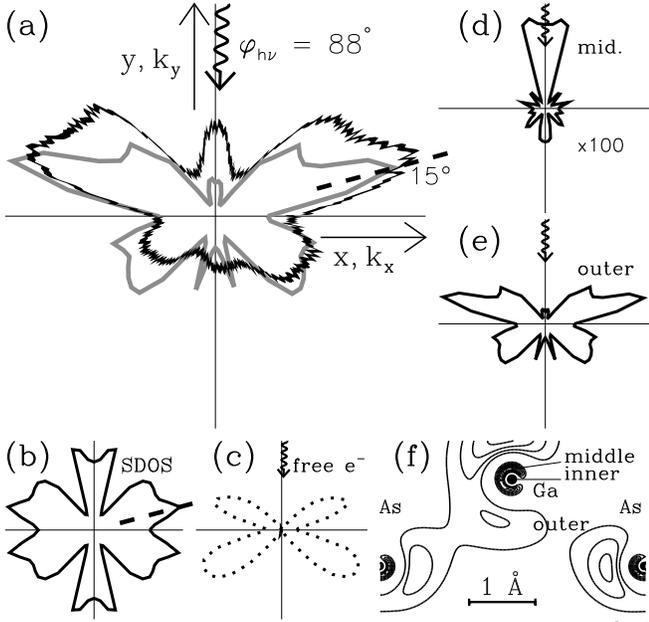}}
\caption{
Photocurrent at $-4.0$ eV and $k_{\parallel} = 0.6$ {\AA}$^{-1}$ 
($\vartheta = 19.4^{\circ}$), 
SDOS and charge density for C$_{2}$+A$_{3}$ states: 
(a) measured (black) and calculated (grey) current for 
 light incident from $88^{\circ}$; 
(b) SDOS; 
(c) current with a free-electron final state; 
(d)+(e) partial currents emerging from the
  middle (d), and the outer (e) area for light incident from 
  $90^{\circ}$; 
(f) contour plot of the charge density in a plane through the 
  the uppermost As and Ga atoms, at $-4.0$ eV, and for 
  ${\bf k_{\parallel}}$ belonging to the upper right lobe as 
  indicated by the dashed lines in (a) and (b). 
The current in (d) is magnified by $100$ relative to that in (e).
The middle region is the spherical volume between radii 
of $0.08$ {\AA} and $0.29$ {\AA} around the As cores and 
$0.09$ {\AA} and $0.3$ {\AA} around Ga as depicted in (f). 
The outer region is the space outside these spheres.
}
\label{fig:3}
\end{figure}

For this A$_{5}$ state, a comparison of 
the photocurrents in Figs.\ \ref{fig:2} (a)--(d) with the SDOS in 
Fig.\ \ref{fig:2} (e) shows 
that the SDOS controls the overall structure in the current. 
The number and the azimuthal
positions of the lobes are the same for the SDOS and the current, but 
differences occur in intensities and in smaller structures. There are 
$4$ broad lobes in the SDOS and the current, but instead of $8$ maxima 
as for the 
SDOS on a fine scale, the current displays only $4$ maxima.
Figs.\ \ref{fig:3} (a) and \ref{fig:3} (b) for the C$_{2}$+A$_{3}$ 
state show the same overall correspondence and differences in intensities.
The SDOS used here is the partial density of 
states of the orbitals from the uppermost atomic layer. 
The power of the SDOS 
for accurate investigations of the {\em energetic} 
structure is demonstrated with the emission from the A$_{5}$ state.
Here, this excellent agreement between experimental and theoretical 
currents could only be achieved with a small shift in the corresponding 
binding energies by $0.15$ eV, but the effect of this energy 
shift is remarkable, as shown in Fig.\ \ref{fig:2} (d). 
The main reason for this difference 
should be attributed to the calculated surface band structure, 
though the 
inaccuracy is still in the usual range of common theoretical 
uncertainty\cite{gaasband}. The strong 
variation of the shape in Fig.\ \ref{fig:2} (d) reflects the changes
in the SDOS in Fig.\ \ref{fig:2} (e). This is obviously caused by
the flat dispersion of the band which is shown in Fig.\ \ref{fig:1}. 
Thus the number, position and intensity of the 
lobes in the current and SDOS depend very sensitively on details of 
the band structure. Small shifts in the energy cause not only 
shifts in the accompanying peaks, but can give rise to strong 
changes of the entire pattern. This allows an accurate 
determination of the band structure by comparing experimental with 
theoretical scans. The usual simpler approach of identifying the maxima in 
the current as band positions, 
would fail here to give even the correct number: 
For the A$_{5}$ state the band is hit $8$ times 
(cf.\ inset in Fig.\ \ref{fig:1}), which corresponds to the $8$ maxima in
the SDOS,
but the current in Fig.\ \ref{fig:2} has only $4$  maxima, 
and one would overlook half of the band positions.

In contrast to the electronic structure, inversion 
symmetry is lacking in the photocurrent. To understand this, 
we have to consider in addition to the SDOS also the wave 
function $\langle {\bf r} |\Psi_{in}\rangle$ of the initial state,
which are calculated in a LCAO basis. 
For the dangling-bond A$_{5}$ band, the asymmetry between the intensities 
in positive and negative $y$ directions corresponds to the direction 
of the bond. The dangling bond points along the negative 
$y$ direction and into the vacuum in the $z$ direction, as shown in 
Fig.\ \ref{fig:2} (f) by the contour plot of a As-charge density. 
Since the density is located above the uppermost As atoms, there 
is only weak potential scattering for a major part of the excited 
electrons. Therefore, the SDOS and the initial state wave function 
dominate the angular distribution, and a free final plane wave  
may be a sufficient description. The dotted current in 
Fig.\ \ref{fig:2} (d) shows how well this approximation does in 
reproducing the general asymmetry. Because a Fourier transformation 
of a spherical harmonic reproduces the same 
spherical harmonic and
because of the simple spatial and angular momentum structure 
of the dangling-bond state, 
the asymmetry of the current can be identified here with 
that of the charge distribution. Such a connection has 
already been assumed in a former ARUPS analysis\cite{phogeo}, 
and it is quantitatively proven here. 
As a result, the photoemission intensities reflect directly 
the energetic and spatial density of the initial A$_{5}$ state. 
However, this simplicity is not a general rule, as we will 
illustrate for our C$_{2}$+A$_{3}$ case at $-4.0$ eV.

To achieve further insight into the influence of the initial state, 
the importance of different parts of the emitting volume 
is investigated in a novel way. Apart from the charge density maxima in 
the bonding region the initial wave functions also have charge 
peaks closer to the cores, as shown in Fig.\ \ref{fig:2} (f) and 
Fig.\ \ref{fig:3} (f). In the present case, three areas are 
separated by almost spherical nodal surfaces. The separate 
contributions to the current from each of these regions are 
calculated by setting the initial wave function to zero in the 
other two areas. 
Contributions for the emissions from 
$-4.0$ eV are shown in Figs. \ref{fig:3} (d)+(e). From the 
localized middle region arises less than $1 \%$ of the total 
current for both studied binding energies. Contributions from the 
innermost area are even smaller and completely negligible. With a 
free final state for simplicity 
we have repeated these calculations for kinetic 
energies up to $1500$ eV. Below $100$ eV only contributions from 
the outer volume are notable, but at $200$ eV the current from the 
middle area becomes appreciable. This middle area dominates 
above $500$ eV, and 
constitutes the current above $1000$ eV. The spatial scale detected 
by photoelectrons may be discussed in 
close analogy to the energy dependence of common scattering processes
where high wave vector Fourier components probe the more rapidly 
varying wave functions near the core. 
Usually with increasing energies shrinking vicinities of the cores are 
studied. 

At $-4.0$ eV, neither the number nor the position of the 
lobes nor the asymmetry in the intensities is reproduced with 
a final plane wave, as shown in Fig.\ \ref{fig:3} (c). 
Only the inclusion of scattering provides the asymmetry in the 
intensities, as the excellent agreement in Fig.\ \ref{fig:3} (a) 
between experiment and theory shows. Therefore, the simple argument
with the Fourier transformation as in the A$_{5}$ case 
does not work. We trace the origin 
of the intensities into the details of the calculation. Although the 
A$_{3}$ state extends over 4 atomic layers, $80 \%$ of the 
electrons are excited from the density around the
uppermost As atom and from the bonds towards the next Ga atoms. 
The inversion asymmetry of the main peak at $15^{\circ}$ is caused  
by these bonds, whose unsymmetrical charge distribution is shown in 
Fig.\ \ref{fig:3} (f). This lack of the band-structure symmetry 
in the intensities generally 
seems to give a strong hint as to the asymmetry of the charge density.

The dependence of the spatial distribution 
on kinetic energy gives an additional feature of ARUPS 
from valence states and its relation to XPD. The importance of smaller
spatial scales with increasing energy might be understood as a
localization of an effective emitting source. At high 
energies this coincides with the observation that the
XPD pattern from valence bands are nearly identical to those from
the localized core states\cite{valencePD}. The common spatial origin of the 
excited electrons leads to similar currents since in the one-step model the 
final states are exactly the same. It should be noted that at these high 
energies the final state scattering is known to dominate over further
details of the source wave\cite{valencePD,PD}, and especially initial state 
interferences vanish on the average by the finite resolution\cite{average}.
Contrarily, at low kinetic energies valence states may contribute from regions 
where the localized core states vanish. These delocalized regions are the
most important for both studied cases. Whereas with the effective 
localization at the core the XPD patterns reflect the geometry, ARUPS probes
the bonding region with an intensity distribution showing information
about the bonds. This is a new aspect for the interpretation of ARUPS data.

We have presented a joint treatment of the spectral
and spatial features of ARUPS from the model surface 
GaAs(110). The azimuthal scans
reflect in number and angular positions of the lobes the electronic
structure given by the SDOS. By comparing measured and calculated 
currents, huge changes induced by dispersion allow an 
accurate determination of the SDOS. Even for unpolarized light 
there are strong additional intensity modulations, which are 
connected to the initial wave function. 
For an interpretation we adopted 
the XPD picture genralized to delocalized valence states. The 
amplitude of the initial state appears as the local emissivity for 
the spatially distributed source of the electrons to be scattered. 
Photoelectron spectroscopy is sensitive to different 
spatial parts of the initial state, depending on the kinetic 
energy. Ultraviolet photoemission detects the wave function 
in the bonding region outside the core, and the origin of 
photoelectrons are traced for the first time into single bonds. This opens new 
possibilities in the application and interpretation of ARUPS.

This work was supported in part by the Bundesministerium f{\"u}r Bildung, 
Wissenschaft, Forschung und Technologie and by the Materials Sciences 
Division of the U.S.\ Department of Energy, Contract No.\ DE-AC03-76SF0009.

\end{document}